\documentclass[prb,nofootinbib,twocolumn,superscriptaddress]{revtex4} 

\usepackage{graphicx}
\usepackage{dcolumn}
\usepackage{bm}
\usepackage{threeparttable}
\usepackage{times}
\usepackage{mathptmx}
\usepackage{lscape}
\usepackage{natbib}
\usepackage{amsmath}
\usepackage{amssymb}
\usepackage{braket}
\usepackage{comment}
\usepackage{color}

\def\degree{\kern-.2em\r{}\kern-.3em}

\begin{document}


\title{ Microscopic Geometry Characterizes Structure/Potential-Energy Correspondence in a Thermodynamic System  }

\author{Koretaka Yuge}
\affiliation{
Department of Materials Science and Engineering,  Kyoto University, Sakyo, Kyoto 606-8501, Japan\\
}%

\begin{abstract}
{ 
Potential energy landscape (PEL) is essential to determine phase stability, reaction path, and other important physical as well as chemical properties. 
Whereas given PEL can reasonably determine the properties in thermodynamically equilibrium state, 
it is generally unclear whether a set of known property can uniquely and/or stably determines PEL, i.e., understandings of property/PEL correspondence is basically \textit{unidirectional} in the current statistical mechanics. 
Here we make significant advance toward \textit{bidirectional} bridging of this gap for classical discrete systems under many-body interactions. 
Our idea is to focus on characteristic microscopic geometry in configuration space for an exactly solvable system, resulting in a new, important  quantity of "harmonicity in the structural degree of freedom". 
This quantity reasonablly characterizes which structures in equilibrium state have practically unique and stable correspondence to PEL, without requiring any thermodynamic information such as energy or temperature.
The present findings will open a gate to constructing reliable PEL, where its predictive uncertainty can be \textit{a priori} known. 
A significant role of microscopic geometry for \textit{non-interacting} system should  be re-emphasized in statistical mechanics.
}
\end{abstract}



\maketitle


\section{Introduction}
One of the main goal for statistical mechanics is to elucidate relationships between many-body interactions of the system and macroscopic properties in equilibrium and/or non-equilibrium states.
Particularly, potential energy landscape (PEL) plays central role to describe phase stability, reaction path and other important physical and chemical properties matters,  
whereas full knowledge of PEL is typically difficult to obtain. 
So far, many different methods have been developed based on \textit{ab initio} calculations to construct effective (classical) Hamiltonian for the PEL.\cite{nat1,nat2} 
These are direct and powerful when (i) all contributions to determine PEL can be accurately included, and (ii) used calculation data has sufficient information to describe PEL on high-dimensional space, which is not known \textit{a priori}. 
For instance, full  inclusion of temperature-dependent interactions (e.g., lattice vibrational, magnetic and electronic entropy effects) for condensed matters is not practical, especially dimension of exploring space exponentially increases due to increase of the number of components: 
Such effects are typically either approximated or ignored when predicting physical properties of equilibrium states, which sometimes fails to explain and interpret measured properties.\cite{mat1, mat2, mat3, lp,lp2} 
For these cases, it naturally leads to take another stragegy to construct PEL. 
The so-called inverse methods, which basically optimizes PEL to reproduce measured structure and/or other properties, have also been amply developed and applied to discrete (e.g. crystalline solids) or continuous (e.g. liquid) systems.\cite{pe, invmc1, invmc2, invmc3, invmc4, invmc5, inv1, inv2} 
However, both direct and inverse methods always involve an essential problem to construct reliable PEL: 
Generally, the transformation from PEL to structure is a well-posed direct problem, whereas that from structure to PEL is a typical inverse problem: Although different methods have been proposed to characterize PEL for system-specific quantities, it is generally unclear whether the constructed PEL uniquely and/or stably determines a set of known macroscopic property.\cite{many1,many2,many3,many4,many-rej, cpel1, cpel2} 
Up to now, for a modest classical system under pairwise additive interactions, it has been shown that the PE can be determined uniquely from the structural information contained in pair-correlation functions.\cite{pe} However, for systems that involve many-body interactions, it remains an open question whether using such an inverse methods is justified.\cite{many1,many2,many3} 
Especially, for crystalline solids considered as classical many-body system, while it has been typically assumed that pair correlations up to a sufficiently large interatomic distance can uniquely (or almost uniquely) determine higher-order correlations, this assumption has been rejected theoretically.\cite{many-rej} 
Thus, understanding of the correspondence between structure in equilibrium state and PEL  (hereinafter called SPE correspondence) is \textit{unidirectional} in the current statistical mechanics.
Based on classical statistical mechanics, we here make significant advances in \textit{bidirectional} bridging of the gap for a broad class of classical systems, under many-body interactions with multiple components. We find a quantity to universally characterize which microscopic structures have unique and stable correspondence with PEL. 
We find that these structures can be known \textit{a priori} without any information about energy or temperature, which will lead to a new approach of constructing reliable PEL. The details are given below. 



\section{Derivation and Concepts}
Herein, we consider a typical classical system with a fixed number of constituents, and whose microscopic structure is specified by a finite number of 
generalised coordinates $\left\{q_1,\cdots , q_f\right\}$. 
The expectation value of a chosen coordination $g$ is then given by the thermodynamic (so-called canonical) average, namely
\begin{eqnarray}
\label{eq:can}
Q_g\left(\beta\right) = Z^{-1}\sum_d q_g^{\left(d\right)} \exp\left(-\beta U_p^{\left(d\right)}\right), 
\end{eqnarray}
where $Z=\sum_d \exp\left(-\beta U_p^{\left(d\right)}\right)$ denotes the partition function with $\beta = \left(k_{\textrm{B}}T\right)^{-1}$, and  the summation is taken over all possible microscopic states in the configuration space. 
If we interpret the thermodynamic average as a map, $\phi_{\textrm{th}}\left(\beta\right)$, acting on the potential energy, Eq.~(\ref{eq:can}) for all possible coordinates $\left\{q_1,\cdots , q_f\right\}$ reads
\begin{eqnarray}
\phi_{\textrm{th}}\left(\beta\right) \cdot \mathbf{U}_p =  \mathbf{Q}\left(\beta\right),\nonumber \\
\quad
\end{eqnarray}
where $\mathbf{U}_p =  \left\{\Braket{U_p|q_1}\cdots,\Braket{U_p|q_f}\right\}$ and $\mathbf{Q}\left(\beta\right)=\left\{Q_1\left(\beta\right),\cdots, Q_f\left(\beta\right)\right\}$ (see Appendix). 
$\mathbf{U}_p$ is a vector representation of potential energy under given coordination, corresponding to specifying PEL. 
When predicting the PEL from a thermodynamically equilibrium structure, the problem is that it is generally unclear whether the inverse map, $\phi^{-1}_{\textrm{th}}\left(\beta\right)$, exists.

Very recently, we have shown\cite{em1,em2,em3} that the map $\phi_{\textrm{th}}\left(\beta\right)$ corresponding to the  thermodynamic average can be approximated in explicit matrix form as $\bm{\Gamma}\left(\beta\right)$ (see Appendix).
We have proven\cite{em2} that the map $\bm{\Gamma}\left(\beta\right)$ becomes exactly identical to $\phi_{\textrm{th}}\left(\beta\right)$ for any given potential energy and for any temperature $\beta$, when the density of microscopic states in configuration space, before applying many-body interactions to the system, is represented by a multidimensional Gaussian distribution. Herein, such an ideal system is defined as a "harmonic system".  

To address the uniqueness of SPE correspondence, we firstly consider whether the inverse map $\bm{\Gamma}^{-1}$ exists. For a harmonic system, we can show  that $\bm{\Gamma}$ becomes exactly (see Appendix)
\begin{eqnarray}
\label{eq:S}
\Gamma_{ik}\left(\beta\right) = -\beta S_{ik},
\end{eqnarray}
where $S_{ik}$ corresponds to the elements of the covariance matrix $\mathbf{S}$ of the density of states (DOS).
Because the eigenvalue of the covariance matrix is positive-semidefinite with non-zero variance for any transformation of coordinates, 
$\bm{\Gamma}^{-1}$ always exists.
This implies that for any harmonic system, we can always write
\begin{eqnarray}
\mathbf{U}_p = \phi_{\textrm{th}}^{-1}\left(\beta\right) \cdot \mathbf{Q}\left(\beta\right),
\end{eqnarray}
for any given PE, i.e. the SPE correspondence is unique.
We have confirmed\cite{em3} that for a broad class of practical classical systems (e.g., for solids on representative lattices, including fcc, bcc, diamond, square and trianglar, and for liquids in a rigid box), the DOS  before including system interactions becomes almost identical to a multidimensional Gaussian as the system becomes large. 
We showed this not only by comparing its landscape with an ideal Gaussian but also by demonstrating statistical interdependence based on random matrix theory with a Gaussian orthogonal ensemble.\cite{em3} 
This indicates strongly that slight deviations of the actual DOS from a multidimensional Gaussian cause changes in the uniqueness of the SPE correspondence. 
In other words, many classical systems can be interpreted as \textit{perturbed} in that they differ slightly from an ideal harmonic system but have the same covariance matrix $\mathbf{S}$. 

\begin{figure}
\begin{center}
\includegraphics[width=0.85\linewidth]
{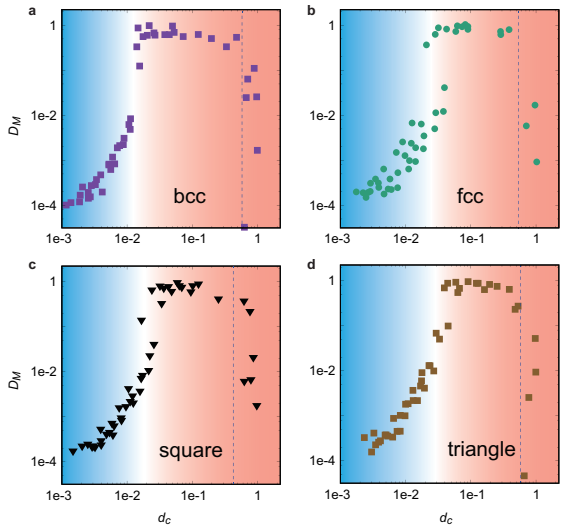}
\caption{Anharmonicity in the structural degree of freedom (S.D.F.) for multiple microscopic states with binary components under different geometries. a: bcc, b: fcc, c: square, and d: triangular lattice.  
The anharmonicity $D_M$ is plotted in terms of its distance $d_c$ (Euclidean metric) from the center of gravity (COG) of the configurational DOS for a non-interacting system. The blue region represents the harmonic regime, while red corresponds to the anharmonic region. Both $D_M$ and $d_c$ are normalised by the maximum value of $d_c$ for the given lattice and composition.}  
\label{fig:dm}
\end{center}
\end{figure}

In order to see how practical system deviates from an ideal harmonic system, we first qualitatively investigate the character of the map $\bm{\Gamma}$ on a two-dimensional periodic lattice with artificial PE provided, which confirm the dependence of difference between $\bm{\Gamma}$ and $\phi_{\textrm{th}}$ on individual microscopic structure. 
The results imply that a certain set of microscopic structures contributes to keep the system harmonic, where distance $d_c$ (Euclidean metric) from the center of gravity (COG) of the configurational DOS for a non-interacting system to the microscopic structure would come into play (see Appendix). In other words, the map $\Gamma$ acts in almost the same way as the thermodynamic average, namely $\Gamma \simeq \phi_{\textrm{th}}$, whereas another set of structures promotes anharmonicity, namely $\Gamma \neq \phi_{\textrm{th}}$. Again, the deviations from harmonicity come from the slight deviations of  the configurational DOS from Gaussian. 
Therefore, if we could estimate quantitatively the harmonic contributions from individual microscopic structures, we can know which observed structures are likely to determine the corresponding PEL uniquely. 
However, the question naturally arises as to how to measure the harmonicity (or anharmonicity). The problem is that (i) the image of the map $\phi_{\textrm{th}}$ depends on both PE and temperature, and (ii) that of $\bm{\Gamma}$ depends on temperature. Ideally, we would like to avoid using any information about the energy or temperature. 
Our solution is based on the fact that the image of the composite map, $\phi_{\textrm{th}}\left(\beta\right)\circ \bm{\Gamma}^{-1}\left(\beta\right)$, is independent of both energy and temperature (see Appendix). 
Therefore, we can determine quantitatively the anharmonicity for any given microscopic structure $\mathbf{Q}_M$:
\begin{eqnarray}
\label{eq:dm}
D_M = d\left(\mathbf{Q}_M, \left(\phi_{\textrm{th}}\circ \bm{\Gamma}^{-1}\right)\cdot\mathbf{Q}_M\right), 
\end{eqnarray}
where $d\left(\quad,\quad\right)$ denotes the distance function.
Because the energy of the system should be independent of any linear transformation of coordinates, we employ hereinafter the natural choice for $d\left(\quad,\quad\right)$ of the standard Euclidean distance. 
The anharmonicity $D_M$ of Eq.~(\ref{eq:dm}) is thus completely independent of many-body interactions and of temperature, which purely reflects the microscopic geometry of the system. Because  $D_M=0$ is required to satisfy $\phi_{\textrm{th}} = \bm{\Gamma}$, the magnitude of $D_M$ can reflect the difference between the maps $\phi_{\textrm{th}}$ and $\bm{\Gamma}$ for a given structure $\mathbf{Q}_M$, i.e. it represents the anharmonic contribution. 
To illustrate this, we show in Fig.~\ref{fig:dm} the value of $D_{M}$ for microscopic structures of binary components on four different lattices, as functions of $d_c$ . 
We see clearly the general trends in Figs.~\ref{fig:dm}a-d: (i) Structures with low values of $d_c$ have low anharmonicity (blue regions) that changes continuously with $d_c$. (ii) When 
$d_c$ increases, a discontinuous change occurs in $D_M$  (white regions), during the transition from the harmonic to the anharmonic region, where $D_M$ takes its maximum value.   (iii) For structures with sufficiently large $d_c$ (right-hand side, beyond the vertical dotted line in the red region), $D_M$ behaves as a multivalued function. 
Since the quantity $d_c$ is a natural measure of ordering (low $d_c$ corresponds to a disordered structure, whereas high $d_c$ indicate an well-ordered one),\cite{cp} 
trends (i) and (ii) means that partially ordered structure has a high anharmonicity, resulting in $\bm{\Gamma}\neq\phi_{\textrm{th}}$, whereas the disordered structure has a lower anharmonicity with $\bm{\Gamma}\simeq\phi_{\textrm{th}}$. 
Trend (iii) has already been shown in a previous study, where multiple choices of PE can provide candidates for the ground-state structure (e.g., the largest value of $d_c$ in a given direction in configuration space).\cite{cp} 
Because $D_M$ and $d_c$ can be determined without any information about many-body interactions or temperature, we can thus know \textit{a priori} which microscopic structures belong to the harmonic or anharmonic regions from Fig.~\ref{fig:dm}.
\begin{figure}
\begin{center}
\includegraphics[width=0.83\linewidth]
{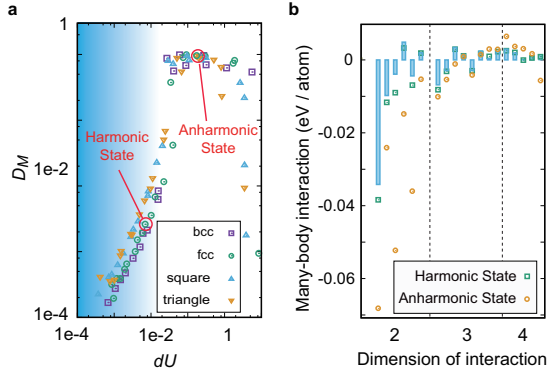}
\caption{Trends in SPE correspondence for binary components under different geometries. a: The relationship between the deviation in PE and the anharmonicity $D_M$. b:  Comparison of the PE in inner product form for the original system (blue bars) and those predicted from the harmonic and anharmonic states. }
\label{fig:dU}
\end{center}
\end{figure}
\begin{figure}
\begin{center}
\includegraphics[width=0.86\linewidth]
{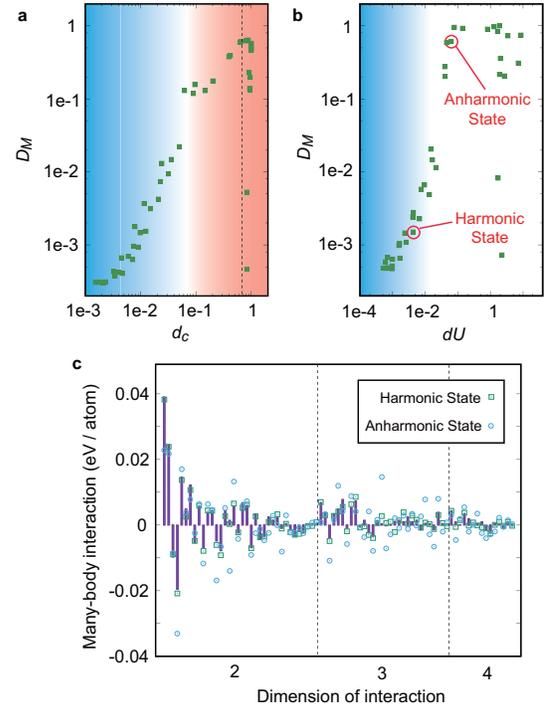}
\caption{Trends in SPE correspondence for quaternary components under fcc lattice. a,b: Value of anharmonicity $D_M$ as a function of $d_c$ and of $dU$. c: Comparison of the PE for the original system (purple bars) and those predicted from the harmonic and anharmonic states in Fig.~\ref{fig:q}b. }
\label{fig:q}
\end{center}
\end{figure}

We then demonstrate how the anharmonicity $D_M$ relates to the actual SPE correspondence. 
We first prepare multiple PEs of binary components on the four lattices with multiple compositions, and apply the thermodynamic average at different temperatures to obtain equilibrium structures, to which the map $\bm{\Gamma}$ is then applied to predict the PEs inversely (see Appendix). The equilibrium structures depend both on the choice of PE ($\mathbf{U}_p$) and on the temperature $\beta$.  The difference between the original and the predicted PEs for each equilibrium structure is measured by 
\begin{eqnarray}
dU =     d\left(\mathbf{U}_p, \left(\bm{\Gamma}^{-1}\left(\beta\right)\circ \phi_{\textrm{th}}\left(\beta\right)\right)\cdot\mathbf{U}_p\right)\cdot\beta.
\end{eqnarray}
Figure~\ref{fig:dU} shows the resultant relationship between the anharmonicity $D_M$ for the constructed structures and the corresponding deviation $dU$ in the PE. 
We  see clearly that when the microscopic structure is in the harmonic region (blue region), $dU$ increases continuously with increasing anharmonicity, while in the anharmonic region, its behaviour becomes complex (i.e. it becomes a multivalued function). This can be expected from the results of Fig.~\ref{fig:dm}, where in the anharmonic region (red region), the deviation of the map $\bm{\Gamma}$ from $\phi_{\textrm{th}}$ is maximised or behaves as a multivalued function. For instance, if we choose a harmonic and anharmonic state constructed from the same PE but for different temperatures (shown in Fig.~\ref{fig:dU}a), we see clearly in Fig.~\ref{fig:dU}b that the PE from an anharmonic state does not reproduce the landscape of the original PE even qualitatively, whereas the PE from the harmonic state (even near the border between the harmonic and anharmonic regions) can reproduce the landscape of the original PE reasonably well. These results demonstrate that microscopic structures in the harmonic state make perturbed contributions to an ideally harmonic system, leading to perturbed changes in the uniqueness and stability of the SPE correspondence. The important point here is that, although we artificially prepared multiple PEs on multiple lattices with different compositions, Fig.~\ref{fig:dU}a shows that for the harmonic region, the map $\bm{\Gamma}^{-1}$ always results in a PE that has a slight deviation $dU$ from the original PE, which exhibits a monotonic increase with increasing anharmonicity. Such a tendency cannot be observed if there are multiple candidate PEs that differ far beyond $dU$. 
Therefore, microscopic structures in the harmonic region can be guaranteed to predict the PE inversely in relation to both uniqueness and stability. Furthermore, the characteristics of the DOS in configuration space also support the uniqueness and stability of the SPE correspondence in the harmonic region. It has been shown\cite{em1} that the DOS near its center of gravity (COG) becomes almost identical to a multidimensional Gaussian, for which deviations generally increase with increasing distance from the COG. Therefore, it is reasonable that contributions from microscopic structures near the COG should be harmonic, i.e. $\bm{\Gamma}\simeq \phi_{\textrm{th}}$.

So far, SPE correspondence can be well-characterized by $D_M$ for binary systems. We finally demonstrate that the above discussions can also be applied when number of components increases. In analogy to Figs.~\ref{fig:dm} and~\ref{fig:dU}, we show in Fig.~\ref{fig:q} the value of $D_M$ as a function of $d_c$ and $dU$ for microscopic structures of quaternary components on fcc lattice. In a similar fashion to binary systems, we can classify which microscopic structures belongs to harnomic or anharmonic region from Fig.~\ref{fig:q} a, which results in a similar relationship between $D_M$ and $dU$ shown in Fig.~\ref{fig:q} b. For instance, when we focus on a harmonic and anharmonic state in Fig.~\ref{fig:q} b, inversely predicted PE for harmonic state (open squares in Fig.~\ref{fig:q} c) correctly captures the landscape of original PE (filled bars), while that for anharmonic state (open circles) fails even qualitatively, which has also been shown in the case of binary system, Fig.~\ref{fig:dU}. 
We therefore have demonstrated that uniqueness and stability of SPE correspondence for classical many-body system can be universally characterized by a newly-introduced concept of \textit{harmonicity in the structural degree of freedom}, which can let us \textit{a priori} know which microscopic structures provides inverse problem from PEL to structure as well-posed in practice.


\section{Conclusion}

In this paper, we theoretically investigated for a classical, many-body, multicomponent system, correspondence between structure and potential energy landscape (SPE corresondence) in terms of the underlying microscopic geometry. We introduced the new quantity of \textit{harmonicity in the structural degree of freedom} derived from an exactly solvable system, thereby  successfully characterising the SPE correspondence. The present results predict the PEL uniquely and stably with residual uncertainty for structure given, without requiring any thermodynamic information about energy or temperature. The present results therefore will open a gate to constructing reliable, temperature-dependent PEL of multicomponent sysmtem with many-body interactions, where existing methods fail to fully include. Especially, inclusion of temperature-dependent contributions, e.g., lattice vibrational, magnetic and electronic entropy, is straightforward (see Appendix).  
We note that a complete understanding of the SPE correspondence will require further study, i.e., (i) how bijection-breaking occurs when $D_M$ increases, and (ii) especially for structures in the anharmonic region, thecorrespondence remains unclear. For the latter, we can see from Fig.~\ref{fig:dU} that several microscopic structures have low anharmonicity but high $dU$, which indicates directly that the SPE correspondence is not unique for such structures. For other structures in the anharmonic state, the correspondence is unclear. Herein, we propose at least a partial set of microscopic structures for which the SPE correspondence is practically guaranteed. As a practical matter, when we predict
the PEL from a given structure, we cannot illustrate simply a relationship like those shown in Figs.~\ref{fig:dU}a, b or Fig.~\ref{fig:q}b because we do not know the original PEL. Alternatively, we can construct relationships between the anharmonicity $D_M$ and $d_c$ for multiple microscopic structures constructed with finite system size, such as those shown in Fig.~\ref{fig:dm} or Fig.~\ref{fig:q}a. From such relationships, we can therefore know \textit{a priori} whether a given structure is in the harmonic or anharmonic region. In the former case, the structure provides $\bm{\Gamma}\left(\beta\right) \simeq \phi_{\textrm{th}}\left(\beta\right)$, which is thus guaranteed to predict the PE inversely, with residual structural uncertainty that corresponds to the anharmonicity $D_M$. 
Therefore, when the given structure is in the harmonic state, the present results can also be applied to confirm the validity of a constructed PEL by existing methods.


\section*{Acknowledgement}
This work was supported by a Grant-in-Aid for Scientific Research (16K06704) from the MEXT of Japan, Research Grant from Hitachi Metals$\cdot$Materials Science Foundation, and Advanced Low Carbon Technology Research and Development Program of the Japan Science and Technology Agency (JST). The author expresses cordial thanks to Prof. Shu Kurokawa and Prof. Yoshikazu Tabata at Kyoto University for fruitful discussions of the present study.

\section*{Appendices}

\appendix
\section*{Description of potential energy for classical systems.}
For classical systems, the potential energy of any microscopic state $r$ of a given system $p$ can be represented completely by
\begin{eqnarray}
\label{eq:u}
U^{\left(r\right)}_p = \sum_{i=1}^f \Braket{U_p|q_i} q_i^{\left(r\right)},
\end{eqnarray}
where $\Braket{\quad|\quad}$ denotes the inner product, i.e. the trace over all possible microscopic states in 
configuration space. For instance, for a discrete system on a periodic lattice, it has been shown\cite{ce} that the generalised Ising model\cite{ce} provides the corresponding complete basis functions $\left\{q_i^{\left(r\right)}\right\}$. 
Equation~(\ref{eq:u}) corresponds to representing the PE in terms of the given coordinates $\left\{q_1,\cdots , q_f\right\}$, in inner product form. 
The present study describes the PE in terms of equation~(\ref{eq:u}).

\section*{Details of map $\bm{\Gamma}$ in matrix form.}
The map $\bm{\Gamma}\left( \beta \right)$ can be given explicitly as follows:\cite{em2}
\begin{eqnarray}
\label{eq:emrs}
&&\mathbf{Q}_{\textrm{ave}} + \bm{\Gamma}\left(\beta\right)\cdot \mathbf{U}_p \simeq \mathbf{Q}\left(\beta\right)  \nonumber \\
 &&\Gamma_{ik}\left(\beta\right) = -\beta \sqrt{\frac{\pi}{2}} \Braket{q_k}_2 \Braket{q_i}_{k+}, 
\end{eqnarray}
where $\mathbf{Q}_{\textrm{ave}}=\left\{\Braket{q_1}_1,\cdots,\Braket{q_f}_1\right\}$ denotes the average structure of a non-interacting system, 
$\Braket{\quad}_1$ and $\Braket{\quad}_2$, respectively, denotes the linear average and standard deviation over all possible microscopic structures. Furthermore, $\Braket{q_i}_{k+}$ denotes the linear average of $q_i$ over all microscopic structures for which the $k$-th coordinate satisfies $q_k \ge \Braket{q_k}_1$. The most important point here is that these standard deviation and averages are calculated for a \textit{non-interacting} system. This implies directly that we can construct the matrix $\bm{\Gamma}$ \textit{a priori} without requiring any information about the many-body interactions.  For simplicity, we represent structure measured from $\mathbf{Q}_{\textrm{ave}}$ throughout the paper, i.e. describing structure as $\mathbf{Q} - \mathbf{Q}_{\textrm{ave}}$.
Because the DOS for a non-interacting system solely reflects the underlying microscopic geometry (e.g., the type of lattice), equation~(\ref{eq:emrs}) strongly indicates the significant role of the geometry in determining the equilibrium structure.

\section*{Derivation of equation~(\ref{eq:S}).} 
Let us introduce the DOS in two-dimensional configuration space for a non-interacting, harmonic system, $g\left(q_i,q_k\right)$, with its $2\times 2$ covariance matrix $\mathbf{M}$. 
Then $\Braket{q_i}_{k+}$ in equation~(\ref{eq:S}) can be given by the following expression:
\begin{eqnarray}
\label{eq:mcov}
\Braket{q_i}_{k+} &=& 2 \int_{-\infty}^{\infty} \int_0^{\infty} q_i\cdot g\left(q_i,q_k\right) dq_k dq_i \nonumber \\
&=& \sqrt{\frac{2}{\pi}} M_{12}\Braket{q_k}_2^{-1} \lim_{q_k\to\infty}\left[1 - \exp\left\{ \frac{q_k^2}{\left(\Braket{q_k}_2\right)^2} \right\}\right] \nonumber \\
&=& \sqrt{\frac{2}{\pi}} M_{12}\Braket{q_k}_2^{-1}, 
\end{eqnarray}
where $M_{12}$ denotes the off-diagonal element of $\mathbf{M}$, i.e. the covariance of the distribution $g\left(q_i,q_k\right)$. 
Substituting equation~(\ref{eq:mcov}) into equation~(\ref{eq:emrs}), we obtain 
\begin{eqnarray}
\Gamma_{ik} = -\beta M_{12}. 
\end{eqnarray}
This directly shows that the matrix $\bm{\Gamma}$ is $-\beta$ times the covariance matrix for the DOS, $g\left(q_1,\cdots, q_f\right)$, i.e. identical to equation~(\ref{eq:S}).

\section*{Example for difference between map $\bm{\Gamma}$ and thermodynamic average $\phi_{\textrm{th}}$ on a periodic lattice.}
To illustrate qualitatively the character of the map $\bm{\Gamma}$ for a practical system, we prepared an artificial PE on a two-dimensional triangular lattice and applied the thermodynamic average $\phi_{\textrm{th}}$ at both low ($\beta_l$) and high ($\beta_h$) temperatures, resulting in partially ordered and well-disordered structures, respectively (see Supplementary). 
The results are summarised in Fig.~\ref{fig:g}.  The PE of the prepared triangular lattice is shown in Fig.~\ref{fig:g}a.  Panels 1a-d  show the results for the low-temperature examples, while panels 1e-h give the corresponding high-temperature results.  Note that the PE distributions in Figs.|~\ref{fig:g}a and ~\ref{fig:g}e are identical. Application of the thermodynamic map $\phi_{\textrm{th}}$ to Fig.~\ref{fig:g}a yields the partially ordered structure shown in Fig.~\ref{fig:g}b, with the corresponding structural parameters shown.  Next, application of the inverse map $\bm{\Gamma}^{-1}$ yields the PE distribution shown in Fig.~\ref{fig:g}c.  Application of the thermodynamic map $\phi_{\textrm{th}}$ to Fig.~\ref{fig:g}c then yields the phase-separated structure and structural parameters shown in Fig.~\ref{fig:g}d.  Figure~\ref{fig:g}e-h gives the corresponding high-temperature results.
\begin{figure}
\begin{center}
\includegraphics[width=1.02\linewidth]
{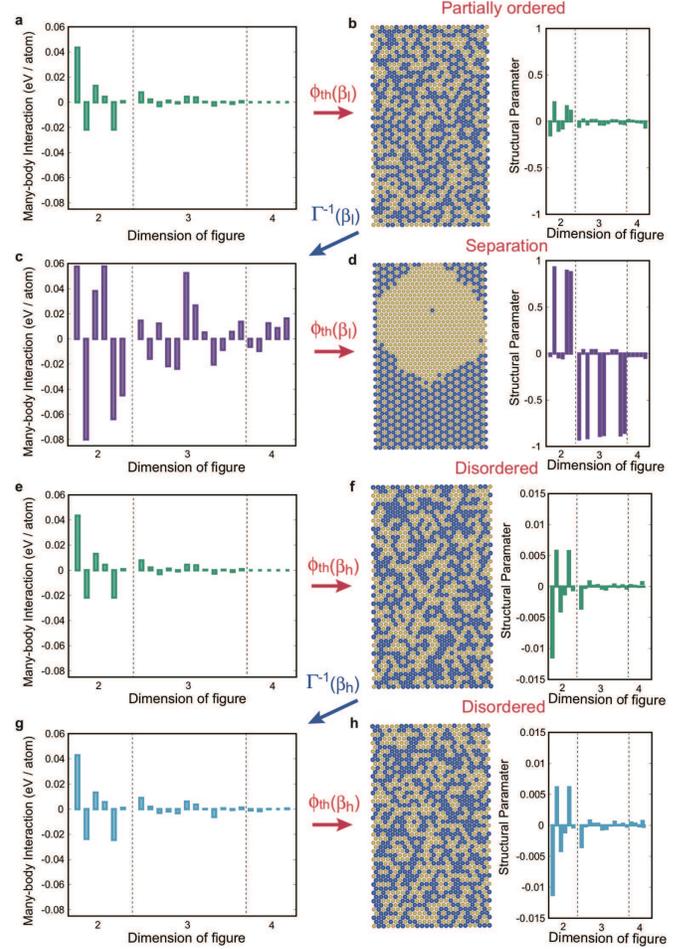}
\caption{Examples of structure/PE correspondence on a two-dimensional triangle lattice. 
Predicted structure in thermodynamic equilibrium state (b, f) from corresponding Potential energy (PE) (a, e). Inverse prediction of PE (c, g) using map $\bm{\Gamma}$ for harmonic system, and resultant equilibrium states (d, h) using the predicted PE. a-d correspond to low-temperature ($\beta_l$) partially ordered structures, e-h to high-temperature ($\beta_h$) well-disordered structure. Note that the PE of a is identical to that of e. 
} 
\label{fig:g}
\end{center}
\end{figure}
We see clearly that when one considers the partially ordered structure (b) at the low temperature $\beta_k$ under the given PE (a), the inverse prediction using $\bm{\Gamma}^{-1}\left( \beta_l \right)$  results in the PE (c) and corresponding equilibrium structure (d), which are completely different from the original ones [(a) and (b)].
Conversely, for the disordered structure (f) at high temperature $\beta_h$ under the same provided PE (e) [identical to (a)], the inverse prediction of $\bm{\Gamma}^{-1}\left(\beta_h\right)$ shows excellent agreement between both the PE (g) and the structure (h) and the respective original ones [(e) and (f)].

\section*{Calculations for Figs.~\ref{fig:dm}, ~\ref{fig:dU}, ~\ref{fig:q} and \ref{fig:g}.} 
To produce Fig.~\ref{fig:g}, we prepared a two-dimensional, triangular lattice consisting of 2304 atoms (i.e. $48\times 48$ times the dimensions of the unit cell).
For Figs.~\ref{fig:g}a and e, we include two-body interactions up to the sixth-nearest neighbours (6NN), twelve 3-body interactions, and five 4-body interactions consisting of up to 4NN pairs, where the multisite figures (i.e. figures consisting of multiple lattice points) in Figs.~\ref{fig:g}  (and Fig.~\ref{fig:dU}b) are described in that order. The thermodynamic average is obtained by applying the artificially-prepared PE to a standard Monte Carlo (MC) simulation, with the Metropolis algorithm\cite{mc1} under the canonical ensemble. The matrix $\Gamma$ is constructed by uniformly sampling the possible microscopic structures based on the MC simulation. To describe quantitatively the microscopic structures on the given lattice, we employed a generalised Ising model in which the occupation of site $i$ by an atom of type A (B) is specified by the so-called spin variable, $\sigma_i=+1$ ($-1$) for binary system, and by A, B, C, D is respectively specified by $\sigma_i=+2,+1,-1,-2$ for quaternary system. The structural parameter for a multisite figure $\alpha$ (including the difference in order of constituent basis functions\cite{ce}) can be obtained by taking the linear average of the spin product taken over the lattice points included in $\alpha$,  
$\psi_{\alpha} = \Braket{\prod_{i\in\alpha}\sigma_i}_{\textrm{lattice}}$. 
Because a set of $\psi_{\alpha}$ for all possible figures on the lattice forms a complete basis,\cite{ce} equation~(\ref{eq:u}) can be rewritten as
\begin{eqnarray}
U^{\left(r\right)} = \sum_{\alpha} \Braket{U|\psi_{\alpha}} \psi_\alpha^{\left(r\right)},
\end{eqnarray}
where the many-body interactions used in computing Figs.~\ref{fig:g}, ~\ref{fig:dU} and ~\ref{fig:q} correspond to the inner product, $\Braket{U|\psi_{\alpha}}$. 
Dimension of interaction denotes the number of lattice points that forms individual multisite figures: i.e. for pairs, the dimension is two, whereas for triplets, it is three. 
In Fig.~\ref{fig:dm}, we consider four lattices (bcc, fcc, square, and triangular) with different binary compositions A$_x$B$_{\left(1-x\right)}$ ($x=0.5, 0.25$). To obtain the quantities $D_M$ and $d_c$ defined by equation (5), we describe the microscopic structure quantitatively, including up to 6NN pairs on each lattice. Explicitly, for structure $\mathbf{Q}_M$, $d_c$ is given by $d\left(\mathbf{Q}_M, \mathbf{Q}_{\textrm{ave}}\right)$. 
For the binary systems on bcc, fcc, square, and triangular lattices, respectively, we also included 8, 12, 7 and 12 triplets, as well as 9, 5, 10 and 5 quartets consisting of up to 4NN pairs. Under these conditions, we randomly prepared 24 microscopic structures for each composition $x$ on each lattice (totalling 192 structures) so that the computed values of $d_c$ for the structures ranged uniformly from near the COG of the DOS for the given structures to candidates near and/or at previously known ground-state structures .\cite{cp} In a similar fashion, for quaternary system, we included 36 pair interactions (corresponding to including up to 6NN pair), 30 triplet, and 15 quartet interactions. Then we randomly prepared 48 microscopic structures at equiatomic composition where value of $d_c$ ranges from disordered to ordered structures. To obtain Figs.~\ref{fig:dU} and~\ref{fig:q}, we prepared multiple sets of many-body interactions for the individual lattices used in Fig.~\ref{fig:dm}, with compositions $x=0.25$ and $0.5$ for binary and with equiatomic composition for quaternary system. They exhibits ordering and a phase-separating tendency when the temperature decreases. These interactions are applied to the MC simulation under the canonical ensemble at eleven different temperatures to compute statistical averages that are used to estimate the anharmonicity in the structural degree of freedom (S.D.F) and in the distance from the COG.

\section*{Derivation: The image of the composite map $\phi_{\textrm{th}}\left(\beta\right)\circ \bm{\Gamma}^{-1}\left(\beta\right)$ is independent of $\beta$.} 
Assume that the temperature $\beta$ changes to $\beta' = c\cdot \beta$, where $c$ is a non-zero finite real number. Then, at $\beta$, the PE for a given structure $\mathbf{Q}$ is given by
\begin{eqnarray}
\mathbf{U}_p = \bm{\Gamma}^{-1}\left(\beta\right) \cdot \mathbf{Q}.
\end{eqnarray}
Meanwhile, at $\beta'$, the corresponding PE becomes
\begin{eqnarray}
\mathbf{U}'_p = \bm{\Gamma}^{-1}\left(\beta'\right) \cdot \mathbf{Q} = c^{-1}\mathbf{U}_p,
\end{eqnarray}
because $\Gamma$ is proportional to $\beta$, as shown in equation (3). 
Applying these PEs to the thermodynamic average $\phi_{\textrm{th}}$, we find that the equilibrium structure for any chosen coordinate $g$ at $\beta$ takes the form
\begin{eqnarray}
Q_g\left(\beta\right) = \left\{ \sum_d \exp\left(-\beta U_p^{\left(d\right)}\right) \right\}^{-1} \cdot \sum_d q_g^{\left(d\right)} \exp\left(-\beta U_p^{\left(d\right)}\right); \nonumber \\
\quad
\end{eqnarray}
while that at $\beta'$ becomes
\begin{eqnarray}
Q_g\left(\beta'\right) &=& \left\{ \sum_d \exp\left(-c\beta c^{-1}U_p^{\left(d\right)}\right) \right\}^{-1} \nonumber \\
&&\cdot \sum_d q_g^{\left(d\right)} \exp\left(-c\beta c^{-1}U_p^{\left(d\right)}\right) \nonumber \\
&=& Q_g\left(\beta\right),
\end{eqnarray}
which demonstrates that the image of the composite map $\phi_{\textrm{th}}\left(\beta\right)\circ \bm{\Gamma}^{-1}\left(\beta\right)$ is exactly independent of $\beta$ for any given structure, $\mathbf{Q}$.

\section*{Inclusion of temperature-dependent interactions to PEL in the present ansatz.} 
For crystalline solids, because the lifetime of a particular atomic configuration is typically long enough to achieve dynamical (including lattice vibrational, magnetic and electronic entropy) equilibrium, the partition function can be described by $Z\simeq\sum_d \exp\left\{-\beta\left(U^{\left(d\right)} + F^{\left(d\right)}\left(\beta\right)\right)\right\}$, where $F^{\left(d\right)}\left(\beta\right)$ denotes the free energy for the temperature-dependent dynamical contributions to configuration $d$.\cite{z} If we define $U_F\left(\beta\right)=U + F\left(\beta\right)$ as  the potential free energy (PFE), replacement of PE $U$ by PFE $U_F$ holds true throughout the paper.\cite{em2} Therefore, inverse prediction from measured microscopic structure to PEL by using $\bm{\Gamma}$ automatically includes all possible temperature-dependent contributions.

\end{document}